\documentclass[onecolumn,authoryear]{els-mrw} 
\usepackage{amsmath,amssymb,amsfonts,amsthm,makeidx,graphicx}
\usepackage{txfonts}
\usepackage{helvet}

\usepackage{amssymb}
\usepackage{amsfonts}
\usepackage{mathrsfs}
\usepackage{soul}
\usepackage{color}
\usepackage{hyperref}
\usepackage{bm}
\def\araa{{ARA\&A}}             
\def\apj{{ApJ}}                 
\def\apjl{{ApJ}}                
\def\apjs{{ApJS}}               
\def\apss{{Ap\&SS}}             
\def\aap{{A\&A}}                
\def\baas{{BAAS}}               
\def\mnras{{MNRAS}}             
\def\prd{{Phys.~Rev.~D}}        
\def\prl{{Phys.~Rev.~Lett.}}    
\def\qjras{{QJRAS}}             
\def\nat{{Nature}}              

\newcommand{\jcap}{JCAP}
\newcommand{\YC}[1]{{\color{black} {#1}}}
\definecolor{forestgreen}{rgb}{0.13, 0.55, 0.13}
\newcommand{\mn}[1]{{\color{black} {#1}}}
\begin{document}

\chapter{Cosmology with Cosmic Voids}\label{chap1}

\author[1]{Yan-Chuan Cai}
\author[2]{Mark Neyrinck}

\address[1]{\orgname{Institute for Astronomy, University of Edinburgh}, \orgaddress{Royal Observatory, Blackford Hill \\ Edinburgh,
EH9 3HJ, UK}}
\address[2]{\orgname{Blue Marble Space Institute of Science}, \orgaddress{Seattle, WA, 98104, USA}}

\maketitle

{\bf Glossary} \\
\\
{\bf Alcock-Paczynski effect} a geometrical distortion in observed galaxy statistics arising from an incorrect cosmological model being used to transform redshift to comoving distance \\
{\bf  Baryon Acoustic Oscillations} density correlations on $\sim 150$ Mpc scales from primordial sound waves \\
{\bf  Cosmic Microwave Background} light emitted at the surface of last scattering, $\sim 10^5$ years after the Big Bang, by now redshifted into the microwave range \\
{\bf Integrated Sachs-Wolfe effect} change of photon energy induced by the change of gravitational potentials due to dark energy\\
{\bf Redshift-space distortions} distortions to the spatial pattern of galaxies in a galaxy redshift survey, produced by peculiar motions of galaxies \\
{\bf Sunyav-Zel’dovich effect} CMB temperature shifts from photons scattering off of plasma \\
{$\bm{\omega}$}, the equation of state of dark energy, defined as the pressure divided by density, $\omega =p/ \rho$ \\ 
{\bf Peculiar velocity} the velocity of a galaxy on top of the apparent motion from the Universe's expansion \\
{\bf Dark-matter halo} a gravitationally-bound clump of dark matter, capable of hosting galaxies if massive enough. \\
{\bf Halo mass function} a histogram of dark matter halos of different masses \\
{\bf Two-point statistics} describe the excess probability of finding two galaxies separated by a certain distance relative to a random distribution\\
\\
\\
{\bf Nomenclature} \\
\\
{\bf AP} Alcock-Paczynski effect \\
{\bf BAO} Baryon Acoustic Oscillations \\ 
{\bf BOSS} Baryon Oscillation Spectroscopic Survey (large-volume SDSS survey sampling bright galaxies)\\
{\bf CMB} Cosmic Microwave Background \\
{\bf DES} Dark Energy Survey \url{https://www.darkenergysurvey.org/}\\
{\bf DESI} Dark Energy Spectroscopic Instrument \url{https://www.desi.lbl.gov/}\\
{\bf ISW} Integrated Sachs-Wolfe effect \\ 
{\bf $\bm{\Lambda}$CDM} Our standard cosmological model, with a cosmological constant ($\Lambda$) and cold dark matter\\
{\bf RSD} Redshift-Space Distortions \\
{\bf SDSS} Sloan Digital Sky Survey \url{https://sdss.org} \\
{\bf (t/k)SZ} (thermal/kinetic) Sunyav-Zel’dovich effect \\ 
{\bf VSF} Void Size Function (a histogram of voids with different sizes)\\
{\bf ZOBOV} ZOnes Bordering On Voidness (a watershed-based void finder)\\

\begin{abstract}[Abstract]
Cosmic voids are low-mass-density regions on intergalactic scales. They are where cosmic expansion and acceleration are most dominant, important places to understand and analyze for cosmology. This entry summarises theoretical underpinnings of cosmic voids, and explores several observational aspects, statistics and applications of voids. The density profiles, velocity profiles, evolution history and the abundances of voids are shown to encode information about cosmology, including the sum of neutrino masses and the law of gravity. These properties manifest themselves into a wide range of observables, including the void distribution function, redshift-space distortions, gravitational lensing and their imprints on the cosmic-microwave background. We explain how each of these observables work, and summarise their applications in observations. We also comment on the possible impact of a local void on the interpretations of the expansion of the Universe, and discuss opportunities and challenges for the research subject of cosmic voids.
\end{abstract}

\begin{figure}
   \vspace{-2. cm}
\begin{centering}
    \includegraphics[width=0.7\columnwidth, angle=270]{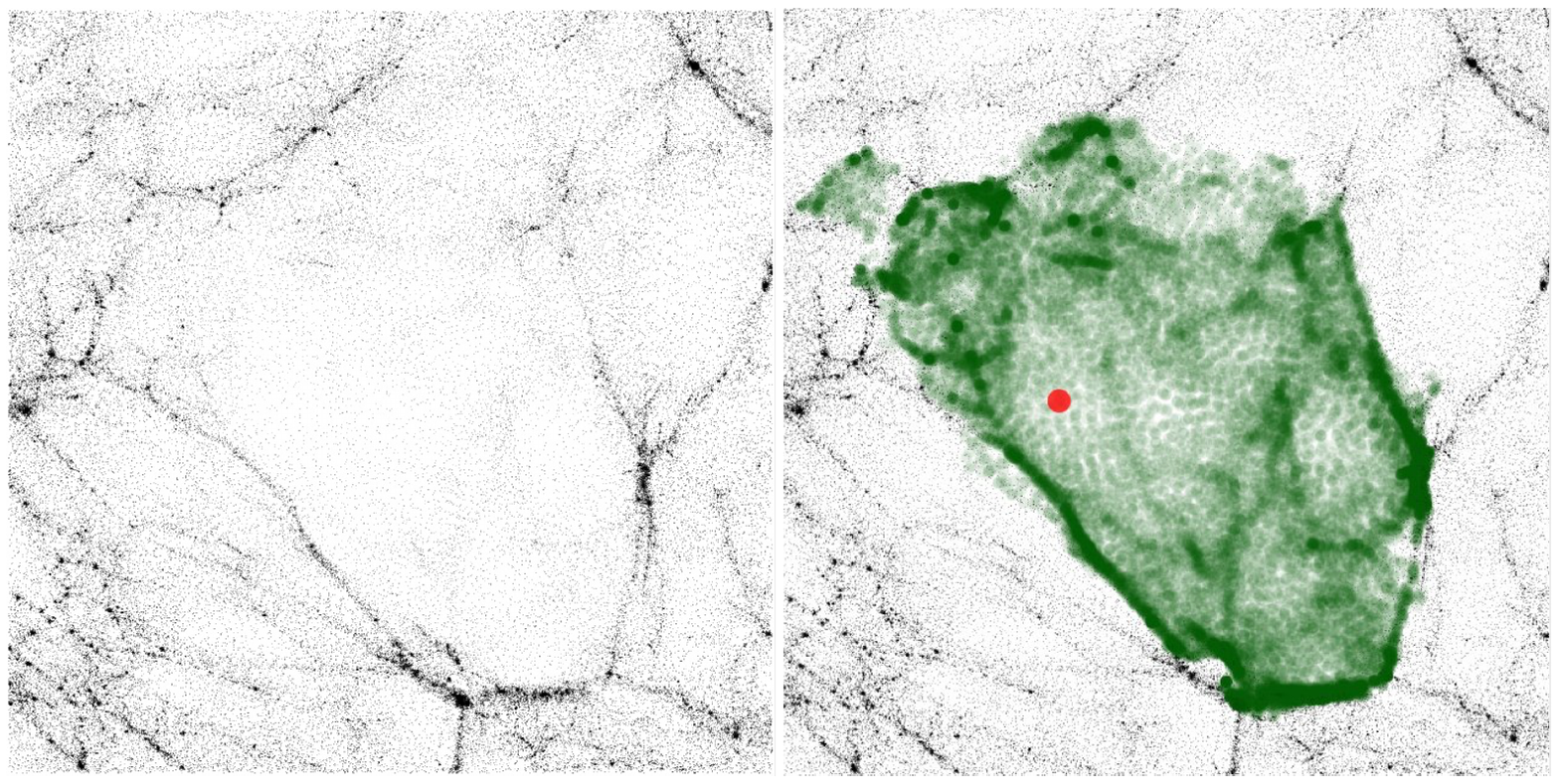}
    \end{centering}
      \vspace{-1.5 cm}
    \caption{Left: The matter in a region of the Millennium simulation \citep{SpringelEtal2005} that contains a large void. Black regions are where matter concentrates, and white regions represent low densities. We can see a large, low-density region near the middle, surrounded by irregularly shaped higher-density boundaries. Colour-coded in green on the right-hand panel is the volume identified as a void by the watershed based void-finding algorithm ZOBOV, the red dot represents the void centre, chosen to be the lowest density point \citep{zobov, Colberg2008}.  The image shows a (40~Mpc/h)$^2$ region with a thickness of 5~Mpc/$h$ from the simulation. This figure is taken from \citep{Colberg2008}.}
\label{fig:ZOBOV}
\end{figure}

\section{The key points about cosmic voids} 

\ \ \ \ \  $\bullet$  Cosmic voids are large under-dense regions found in the large-scale matter distribution of the Universe. The lowest matter densities in the Universe are found there, reaching several orders of magnitude lower than the mean density of the Universe, but not all the way to zero. There are various ways to define voids; depending on the definition, they may be entirely devoid (empty) of galaxies, or simply have a lower galaxy number density than their surroundings. Voids are one of four major components of the large-scale structure of the Universe, which consists of voids, walls, filaments and knots. These are organised in complex web-like structures, called the cosmic web. Figure~\ref{fig:ZOBOV} gives an example of a void embedded in the cosmic web from a cosmological simulation. \\
\\

$\bullet$ The first discover\mn{ies} of cosmic voids went together with the discovery of the cosmic web. Before their discovery in the 1970s, people were aware of the existence of galaxies and galaxy clusters, and that the Universe was expanding, but the spatial distribution of galaxies and galaxy clusters on large scales was largely unknown. In 1978, two independent studies using different surveys of galaxies came to the same conclusion that intergalactic voids exist. \citet{Joeveer1978} and \citet{Gregory1978} concluded that `Unexpected in our results are the large dimensions of holes ranging from some ten to hundreds of megaparsecs' \citep{Joeveer1978}. In the meantime, it was explicitly pointed out that `there are large regions of space with radii $r > $ 20 Mpc where there appear to be no galaxies whatever.' These established the first observational evidence for cosmic voids, and their connection to the cosmic web. \\
\\

$\bullet$ The density within voids is generally of order 10\% of the mean density of the Universe, but voids often contain sub-cosmic webs themselves \YC{\citep[e.g.][]{vandeWeygaert1993, Dubinski1993,Sahni1994, Aragon-Calvo2010,Aragon-Calvo2013,JaberEtal2024}}, with substantial density fluctuations. The effective radius of a void spans from a few Mpc/$h$ to the order of 100 Mpc/$h$. The majority of the volume of the Universe is filled with voids. According to several major void finding algorithms , the total volume occupied by voids accounts for over 60\% of the total volume of the late-time Universe \citep{Cautun2014,Libeskind2018}. Matter around void centres moves away from them.\\

$\bullet$ The statistical properties of voids are dictated by the initial conditions of the Universe. In the standard $\Lambda$CDM model, these are set by the cosmological parameters of the model. More directly than for other parts of the universe, the evolution of cosmic voids is shaped by cosmic expansion and global dynamics. Cosmic voids therefore contain information about cosmology and astrophysics. Among them, the density profiles of voids and volume distribution function can be used to measure the sum of neutrino masses, test theories of gravity, and constrain cosmological parameters. \\

$\bullet$ Cosmic voids are emerging as a major cosmological probe. Analysis of galaxy redshift surveys alone has detected redshift-space distortions \citep{Kaiser1987} and the Alcock-Paczyski effect (AP) \citep{alcockpaczynski1979} around voids. Adding additional data, such as the cosmic microwave background (CMB) emitted soon after the Big Bang, and background galaxies, has allowed detection of gravitational lensing and the integrated-Sachs-Wolfe effect \citep{Sachs1967}. All these have been used to constrain the strength of clustering of matter in the Universe, the rate by which matter perturbations grow, and the matter-energy content of the Universe.

\section{The evolution of individual voids}
\label{sec:Individual_Voids}
\subsection{The formation of voids in a nutshell}
According to the standard model of the Universe, our Universe began in a hot and dense state, with matter almost perfectly uniformly distributed. Matter clumps such as galaxies were not to be found. There were, however, small density fluctuations thought to be seeded from quantum fluctuations during the phase of inflation, which was well within the first second after the Big Bang \citep{Guth1981,Albrecht1982,Linde1983}. The fluctuations seeded from inflation developed into small ripples of low and high densities on the order of 1 part in $10^5$ in the matter distribution when the CMB was emitted. Driven by the universal gravitational attraction of gravity, these fluctuations continued to grow. Matter continuously moves away from underdensities and flows towards overdensities. The underdense regions grew more and more under-dense, and eventually grew to be (nearly) empty of matter, in regions that now call cosmic voids. 

Individual voids have complex and irregular morphologies, such as the one shown in Figure~\ref{fig:ZOBOV}. Voids are often idealized as convex polyhedra produced in a Voronoi tessellation \citep{IckeVdw1991}, which has theoretical justification \citep{Hidding2016,Neyrinck2018}. Physically, voids can be thought of as bubbles in a foam \YC{\citep[e.g.][]{Icke1987, vandeWeygaert1989, vandeWeygaert2003, Aragon-Calvo2023}} with an interior expansion faster than the cosmic mean. Where they run into each other, they produce the cosmic web of walls, filaments and galaxies.

Dark energy, if conceived as a physical substance, is thought to have constant density throughout the Universe, so voids, where there is very little matter, are the most dark-energy-dominant regions of the Universe. There have even been suggestions that the conditions in voids are crucially responsible for the accelerated expansion we see \cite[e.g.][]{Wiltshire2009,Racz2017}. But the mainstream picture is that a sufficiently good approximation to the dynamics and for observations, even in voids, uses Newtonian gravity on top of an expanding background given directly by cosmological parameters \citep[e.g.][]{Kaiser2017}.

\subsection{The spherical expansion model of voids}

\begin{figure}
\vspace{-5.5 cm}
\begin{centering}
    \includegraphics[width=1.0\columnwidth]{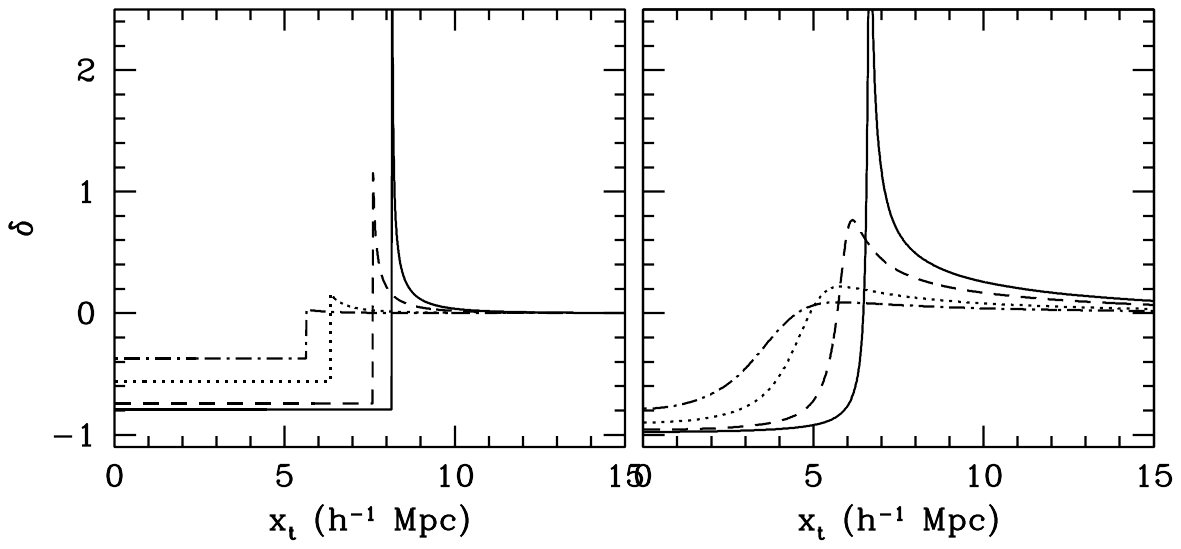}
    \vspace{-9 cm}
\end{centering}    
\caption{The evolution of the void density profile from its initial conditions to shell-crossing, in two idealized cases: that of a spherical top-hat under-density (left), and a void with an averaged profile in the SCDM (standard Cold Dark Matter, with $\Omega_m=1$) model} (BBKS, Eq.~7.10) \citep{BBKS}. Different line styles represent different epochs of the Universe with scale factors $a=0.05$ (dash-dot line), 0.1 (dotted line), 0.2 (dashed line) and 0.3 (solid line). We can see that the voids become emptier as they expand, with matter accumulating at the edge at late epochs, forming high density boundaries. The figure is taken from Figure 3 of \citep{ShethVD}, see also Figure 6 of the review article \citet{vdW2016}. 
\label{Fig:SphericalModel}
\end{figure}

Although voids have irregular shapes, the average of a large sample of them can be approximated by a spherical underdensity. Spherical symmetry is also relevant because a void (at least, its inner density contours) tends to become more spherical as time passes \citep{Icke1984}. A spherical model has proven quite useful to understand their dynamics.

The spherical evolution model was initially introduced to understand the evolution and collapse of high-density regions, or dark matter halos \citep{Bertschinger1985, Blumenthal1992,ShethVD}. But it can be applied to any spherically symmetric system, including voids. The model assumes that a void starts from a 3D spherical underdensity embedded in an infinite background at the mean density. The assumption of spherical symmetry reduces the 3D problem to that of a 1D density profile. Two examples are shown in Figure~\ref{Fig:SphericalModel}. According to Gauss' Law, the acceleration of a spherical matter shell at the radius $r$ is solely governed by the total mass within $r$, and the background dark energy density $\rho_\Lambda$, as given by the following equation:
\begin{eqnarray}
\label{eq1}
\frac{\ddot{r}}{r} &=& -\frac{4\pi G}{3}\left[\rho_{\rm m}+(1+3w)\rho_{\Lambda}\right].
\end{eqnarray}
$\rho_{m}$ is the mean total matter density within $r$; $\rho_{\Lambda}$ is the background dark energy density, $w$ is the equation of state of dark energy, $G$ is the gravitational constant; $\dot r$ represents the time derivative, or velocity, and so $\ddot r$ is the acceleration. The above equation can be solved when the boundary conditions are specified. If the initial density of the void is indeed smaller than the mean density of the Universe, such as a spherical top-hat underdensity within the void radius $r_{\rm v}$, as illustrated in Figure~\ref{Fig:SphericalModel}, the underdensity will expand faster than the Hubble expansion. The matter shells immediately outside $r_{\rm v}$ expand slower than the matter shell at $r_{\rm v}$. This causes matter to build up at the boundary of the void at a later epoch, forming a high-density ridge. The expansion of the inner shell of the void will eventually overtake the outer shell. This is defined as shell crossing for voids. In a matter-dominated Universe with the matter density parameter $\Omega_{\rm m}=1$, this happens when the radius of the underdensity expands by a factor of $1.7$. With mass conservation, we can find that the density contrast of the evolved top-hat void is $\delta=-0.8$. These values have little dependence on cosmology \citep{ShethVD, Demchenko2016}. The mean density of the spherical top-hat void at shell-crossing has been taken as a conventional definition for voids.

With a smoother initial density profile such as the one illustrated at the right-hand panel of Figure~\ref{Fig:SphericalModel}, the evolution of the void density profile is qualitatively similar to the case of the spherical top-hat profile except that the density ridge at the boundary of the void is smoother. The inside of the void is in a sense pristine, with various things preserved from the initial conditions: its initial void density profile is partly preserved, as is even its primordial fluctuation pattern \citep{Aragon-Calvo2013,NeyrinckYang2013,NeyrinckEtal2022} at a later epoch. This in principle allows us to use the observed profile of voids at late times to constrain the initial conditions of the Universe, and therefore measure the cosmological parameters of a model universe.

\begin{figure}
\begin{centering}
    \includegraphics[width=0.48\columnwidth]{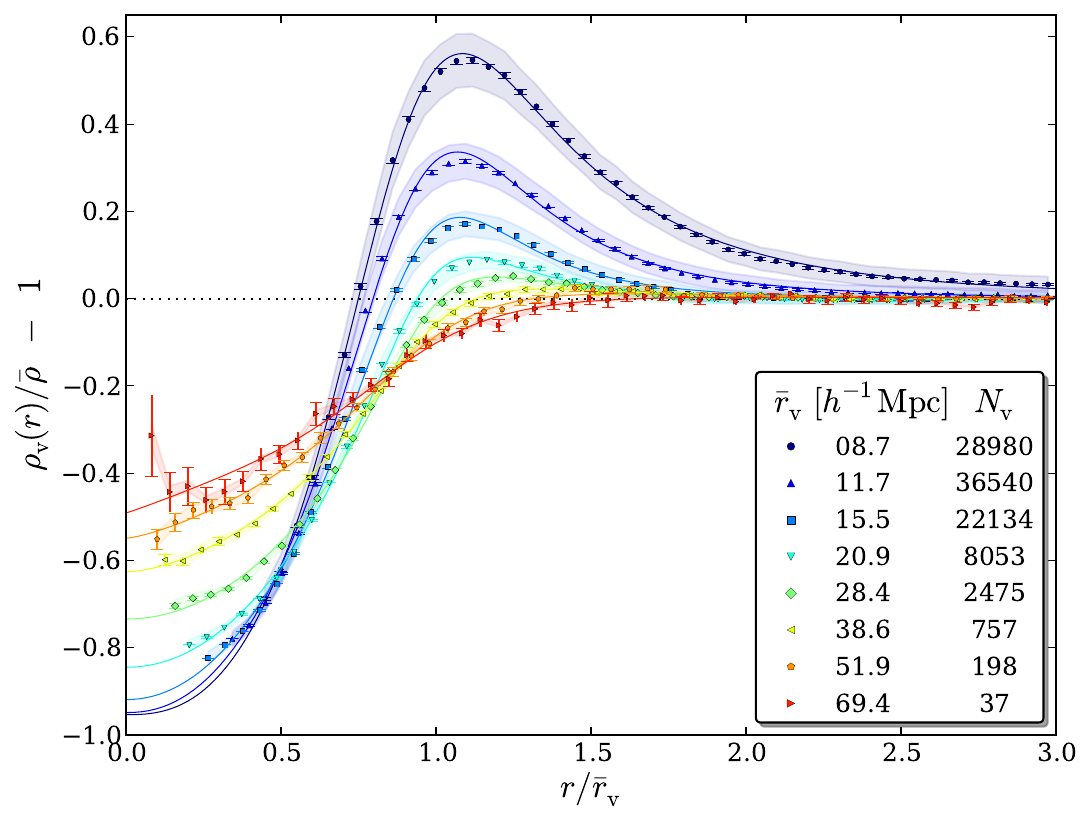}
    \includegraphics[width=0.48\columnwidth]{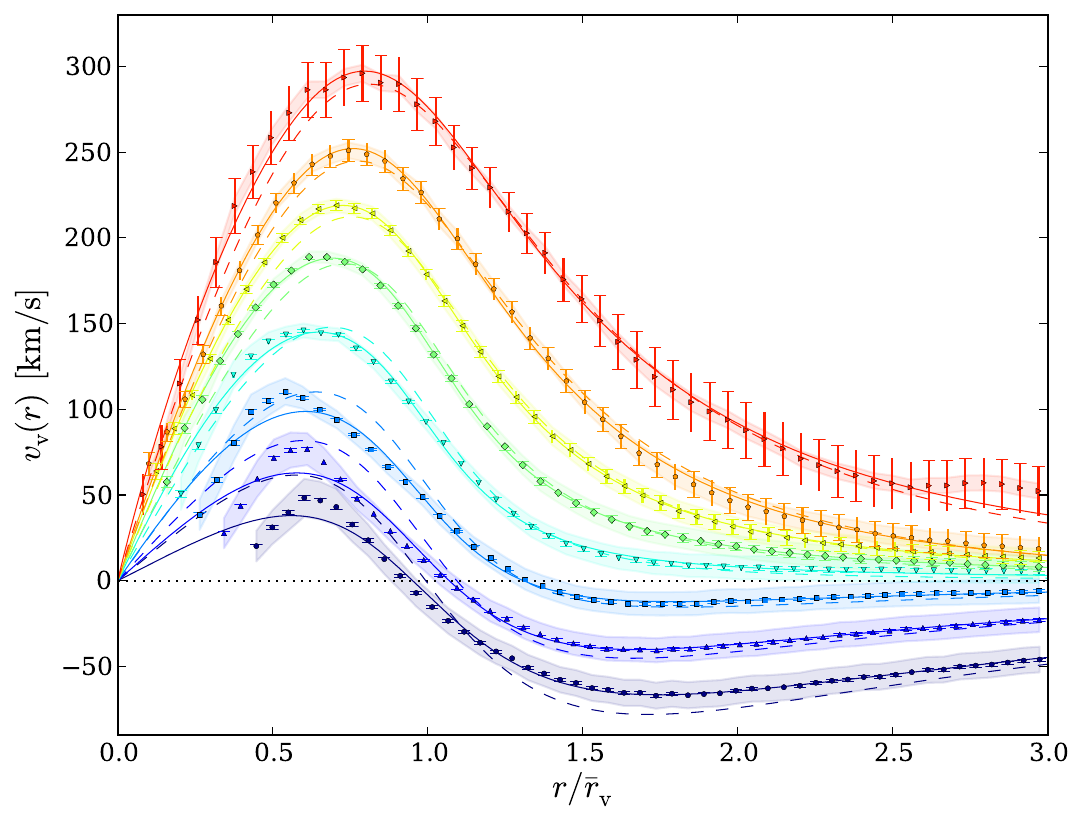}
    \end{centering}
    \caption{The averaged density (left) and velocity (right) profiles of voids from cosmological $N$-body simulations. The averaged size of voids in each bin is represented by their effective radius $\bar{r_v}$, with $N_v$ indicating the number of voids in each bin. We can see that small voids tend to have high and steep density at their edges, while large voids have gentle density boundaries. The interior of voids has positive radial velocities -- they are expanding. At large radii, large voids are also expanding, but small voids are contracting, having negative velocities at large radii. Solid lines represent the best-fit solutions from equation~\ref{Eq:voidprofile} for the density and its corresponding velocity profiles by assuming linear coupling between density and velocity. Dashed lines show the linear theory predictions obtained from evaluating the velocity profile equation at the best-fit parameters obtained from the density stacks. The figure is taken from Figure~1 of \citet{Hamaus2014b}.
}
\label{Fig:VoidProfile}
\end{figure}

Testing against measurements from cosmological simulations, the spherical evolution model has been shown to be a reasonable approximation in tracking the evolution of the density profiles of large voids, but it fails for smaller voids \citep{Demchenko2016}. It was pointed out that it was important to account for the peculiar motions of individual voids for the model \citep{Massara2018}. It was recently shown \citep{Verza2024} that by using an effective moving barrier, adjusting the density field with respect to the void formation threshold, the predicted Lagrangian void density profiles can match up with measurements from cosmological simulations. 

The spherical model and its variants provide us with physical intuitions for the evolution of voids on average -- that voids are expanding relative to the background Universe. They also provide the basis for deriving analytical expressions for the distribution function of the sizes of voids. The shell-crossing definition of a void boundary can be useful beyond the idealised spherical model, with aspherical or even non-convex boundaries \citep{Falck2014}.

\subsection{The density profiles of voids}
There has not been a first-principle derivation for the density profiles of voids, but an empirical fitting formulate with free parameters was found in simulations \citep{Hamaus2014b}. It provides good descriptions for the average profiles of voids at low redshifts (see the left-hand panel of Figure~\ref{Fig:VoidProfile}),
\begin{equation}
\label{Eq:voidprofile}
\delta(r)=\delta_c\frac{1-(r/r_s)^{\alpha}}{1+(r/r_v)^{\beta}},
\end{equation}
where $\delta(r)=\rho(r)/\bar{\rho}-1$ is the density contrast with respect to the background density of the Universe. The parameters are: $\delta_c$ -- the density contrast at the void centre; $r_v$ -- the effective radius of the void; $r_s$ -- a scale radius at which the density of the void equals to the mean density of the Universe; two power-law indices $\alpha$ and $\beta$ that control the inner and outer slopes of the void profile. The formula applies for voids found with the ZOBOV algorithm \citep{zobov}. We can see from Figure~\ref{Fig:VoidProfile} that it fits well the density profiles of the voids with a wide range of void radii. In addition, it was demonstrated that the mean peculiar velocities around these voids follow the predictions from linear theory reasonably well \YC{\citep{Hamaus2014b,Schuster2023}}, \YC{confirming the scenario shown by \citet{vandeWeygaert1993}} i.e., the radial velocity profile $v(r)$ around the void centre, and the density perturbation within the radius $r$, $\bar \delta(r)$, are linearly coupled by constant factors: 
\begin{equation}
\label{eq:LinearVelocity}
v(r)=-\frac{1}{3}r a H(a) \bar \delta(r), 
\end{equation}
where 
\begin{equation}
\label{eq:delta_bar}
\bar \delta(r) = \frac{3}{r^3}\int^r_0\delta(r')r'^2dr'
\end{equation}
(see the right-hand panel of Figure \ref{Fig:VoidProfile}).
For voids with a wide range of radii, they all exhibit positive radial velocities inside the void radius, indicating that voids are expanding relative to the expansion of the Universe. The dynamically simple nature of voids is an advantage of voids compared to high-density regions because they better preserve cosmological information from the initial conditions. This is one reason for the large amount of cosmological information one can obtain using observations of cosmic voids.  

\subsection{Applications}
We can see from the equation of spherical evolution that the strength of gravity (via $G$) and the properties of dark energy (via $w$) affect the evolution of the voids. In addition, neutrinos, which gravitate but were relativistic in the early Universe and have streaming velocities \citep{szalay1976neutrino}, also play a different role in shaping the evolution of voids. The late-time density profiles of voids, which we can observe, carry the imprints of neutrinos. This allows us to use the profiles of voids to test theories of gravity, constrain dark energy and the properties of neutrinos.

For theories of gravity which predict coupling between the strength of gravity with the matter density, such as $f(R)$ gravity \citep{HuSawicki2007}, the impact of modified gravity on structure formation can be stronger in voids. Voids are made emptier by the fifth force in this model compared to their counterparts in general relativity \citep{Cai2015, Perico2019}. A similar effect is found for the nDGP models \citep{Paillas2019}. 

For neutrinos, due to their streaming velocities, they do not cluster as strongly as cold dark matter. In voids, their mass ratio against dark matter may be relatively high, and as the size of voids may be comparable to the scale of free-streaming of massive neutrinos \citep{Zhang2020}, we may expect a stronger impact of neutrinos on the growth of structure around voids. Indeed, studies using cosmological $N$-body simulations show that voids in massive neutrino cosmologies are less empty compared to the case of no neutrinos \citep{Massara2015, Kreisch2019, Schuster2019}. This can potentially be detected with gravitational lensing \citep{Vielzeuf2023}. Neutrinos, although they have a small mass fraction, can be thought of as hot dark matter; similarly to the above, voids in a model with warm instead of cold dark matter are usually shallower density depressions \citep{Yang2015}.

The ellipticity of voids has also been proposed as a probe for the dark energy equation of state \citep{leepark2009,lavauxwandelt2010,bosetal2012,sutteretal2015}.

\section{The distribution function of voids}

\begin{figure}
\hspace{-0.4cm}
\centering
\includegraphics[width=0.9\columnwidth]
{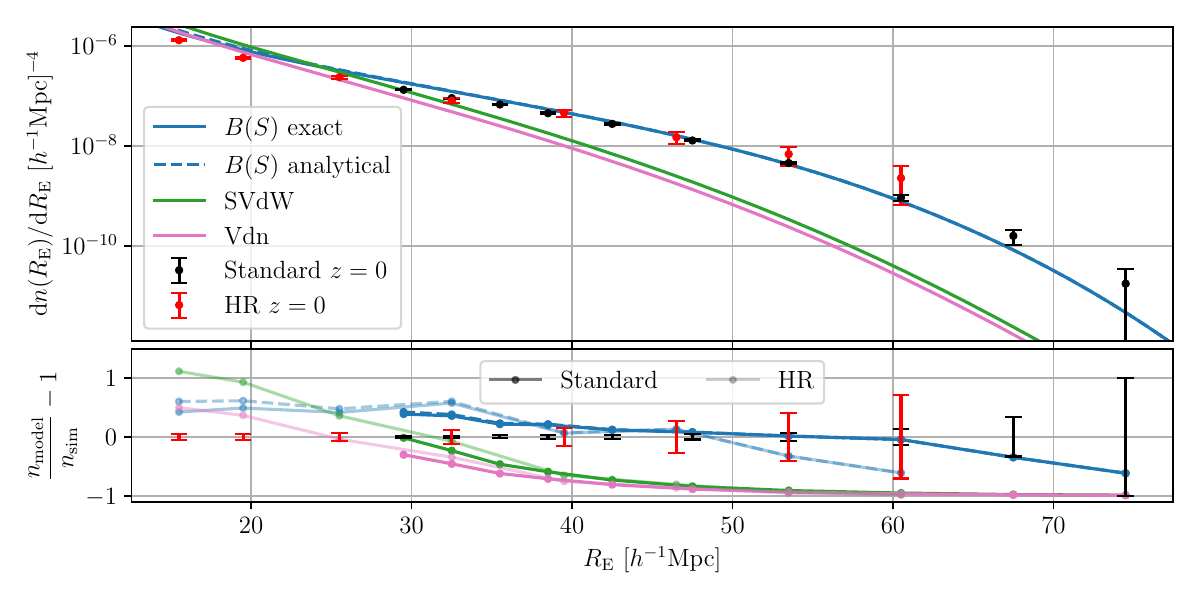}
\caption{Top panel: the distribution function of void sizes (VSF) measured from cosmological $N$-body simulations (black data points with errors and red data points with errors) at $a=1$. The y-axis indicates the number density of voids per unit volume per unit void radius $R_E$ (x-axis). The blue solid and dashed lines show the best-fit model presented in \citet{Verza2024} and its analytical approximations. The model of \citet{ShethVD} (SVdW) and its renormalised version (Vdn) \citep{Jennings2013} are shown in green and pink lines. Bottom panel: relative difference between the
theoretical models in the upper panel with respect to the measurements from simulations. The figure is taken from Figure~4 of \citet{Verza2024}. Here, the Eulerian formation barrier of voids is chosen to be $\delta_v^E =-0.409$, corresponding to the Eulerian linear barrier $\delta_v^{Lin}=-0.623$.} 
\label{Fig:VSF}
\end{figure}

\begin{figure}
\hspace{-0.4cm}
\centering
\includegraphics[width=1.0\columnwidth]
{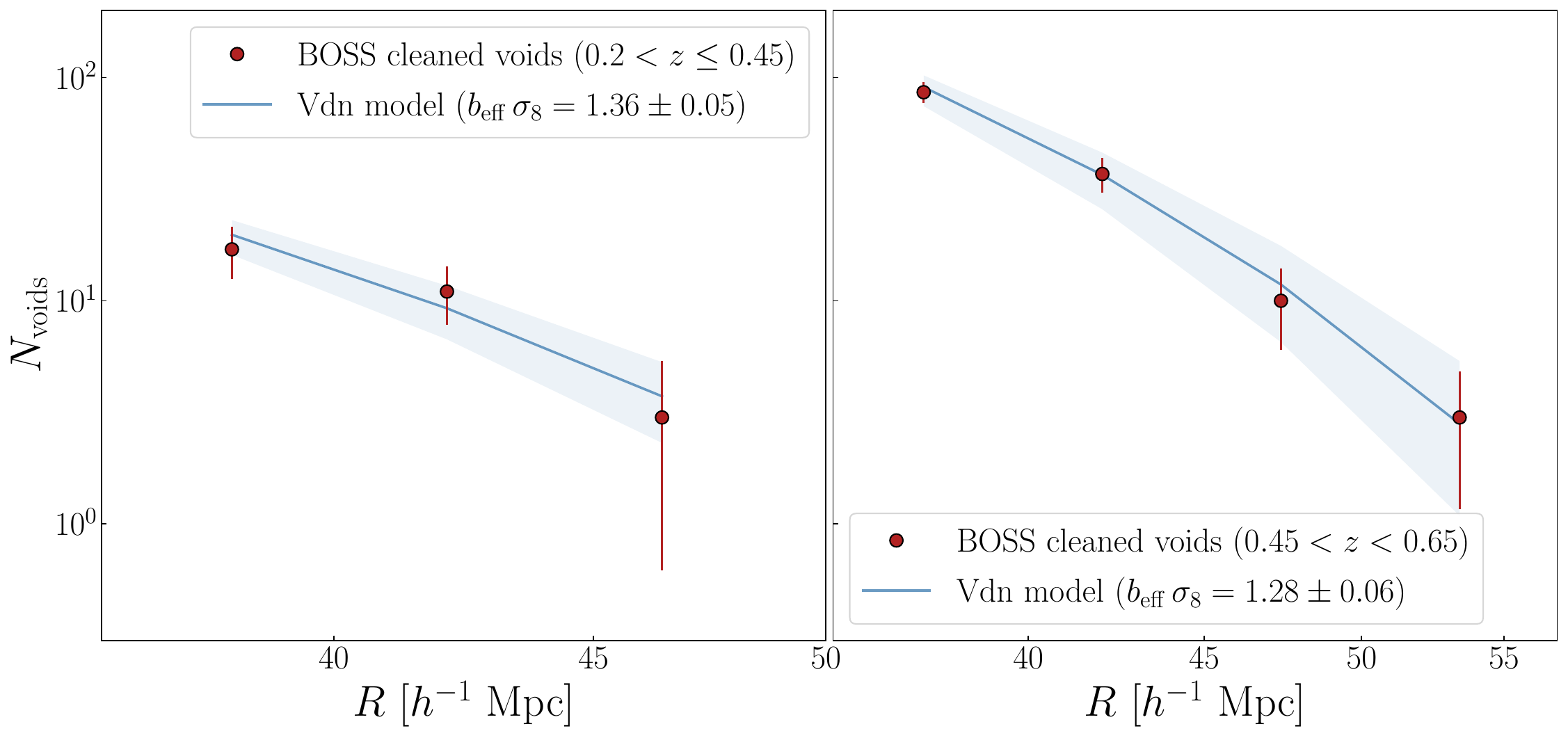}
\caption{Top panel: the size distribution function of voids (VSF) measured with voids found in the SDSS-BOSS-DR12 (Sloan Digital Sky Survey -- Baryon Oscillation Spectroscopic Survey -- Data Release 12) galaxy sample  (red data points with errors) at two different redshift bins shown in the legend. The best-fit model is the Vdn model \citep{Jennings2013}, which is the renormalised SVdW \citep{ShethVD} model, and is extended with two extra nuisance parameters. Priors informed from calibrations with mocks are used for the nuisance parameters in the fit. They are marginalised over for obtaining the order of 10\% measurements for cosmological parameters of matter density $\Omega_m$, amplitude of fluctuations $\sigma_8$ and the dark energy equation of state $w$ \citep{Contarini2023}. The figure is taken from Figure~5 of \citet{Contarini2023}. } 
\label{Fig:VSF_Obs}
\end{figure}

Given a sample of voids, making the histogram of their number counts according to their sizes yields the distribution function of voids, or the VSF. The size of a void is characterised by its volume $V$, or the effective radius, $r_v$ according to $r_v= [3V/(4\pi)]^{1/3}$. Examples of a VSF are shown in \YC{Figures~\ref{Fig:VSF} and \ref{Fig:VSF_Obs}}, where typically small voids are more abundant than large ones. 

In the literature, the distribution function of voids can be named differently. It can be called the void probability function (VPF), or the void abundance function. It was shown in \citet{White1979} that the void probability function, defined in the paper as the probability of a randomly placed region of a given volume being empty, is strongly influenced by correlations of all orders. The VSF therefore includes some cosmological information beyond two-point statistics. It can be used to measure cosmological parameters, including the sum of neutrino masses, and test theories of gravity. For these applications, it is necessary to have theoretical predictions for the VSF.

The VSF is fundamentally connected to the initial conditions and evolution of the individual voids mentioned in the previous section. Analytical predictions for the VSF are usually based on the excursion set theory \citep{Bond1991} in conjunction with the spherical evolution model. Provided with the initial density fluctuations, which are assumed to be Gaussian, the excursion-set theory predicts the probability of initial under-dense regions to form voids in the late-time Universe. The basic idea is that a sequence of density fluctuations $\delta(S)$ with increasing $S$ follows a Brownian random walk, if the filtering of the density field is a Fourier $k$-space top-hat \citep{Bond1991,ShethVD, Zentner2007}. $S=\sigma^2(R_i)$ is the variance of the initial density field smoothed by a top-hat filter of radius $R_i$; in the several-Mpc regime of interest, increasing $S$ means decreasing $R_i$. A random walk that first down-crosses the void formation barrier $\delta_v$ is counted as a void. The probability density of a walk first crossing the barriers between [$\sigma, \sigma+d\sigma$] is expressed as:
\begin{equation}
\frac{df(\sigma, \delta_{v})}{d\ln \sigma}=\frac{\delta_{v}}{\sigma}\sqrt{\frac{2}{\pi}} \exp{\left(-\frac{\delta^2_{v}}{2\sigma^2}\right)}, 
\end{equation}
where $\delta_v=-2.81$ is the linearly extrapolated density to the time of shell-crossing. The corresponding depth of the non-linearly evolved void is $\delta_{NL}=-0.8$. This was first derived in \citep{ShethVD}.

The above excursion set theory was initially developed for predicting the halo mass function. However, applying it to voids leads to violation of volume conservation. This is because voids expand as they evolve. By the time of shell-crossing, its radius would have grown by a factor of 1.7, as mentioned in the previous section, and so their volume would have grown by a factor of $\sim 5$. The sum of the volumes of voids at the late-time Universe exceed their total initial comoving volume. An improved version of the model was derived in \citep{Jennings2013} by imposing volume conservation, which predicts the number density of voids $dn$ per unit logarithmic radius bin $d\ln r_v$ to be:
\begin{equation}
\frac{dn}{d\ln r_v} =\frac{1}{V(r_v)} \frac{df(\sigma, \delta_{v})}{d\ln \sigma} \frac{d \ln \sigma^{-1}}{d \ln r_v},
\end{equation}
where $r_v$ is the late-time evolved radius of voids. A better agreement for the VSFs between simulations and the new model is achieved, after cleaning their simulated void catalogues to retain only non-overlapping voids \citep{Jennings2013, Ronconi2017,Ronconi2019}. 

Once the initial matter density power spectrum is specified, the abundance of voids can be predicted with the above equation. Note that the formation barrier of voids and the corresponding density contrast of the nonlinear void come necessarily from the spherical model, assuming a top-hat initial density. Having different initial density profiles will certainly lead to different formation barriers of voids, and different predictions for the VSF. Nonetheless, this model provides a useful guidance for the general physical picture of voids in the following aspects: the distribution of voids in terms of their sizes follows a Press-Schechter-type \citep{PS} function -- large voids are rare compared to small ones.

Recently, by combining the excursion-set approach with the peak theory formalism in Lagrangian space, \citet{Verza2024} have succeeded in having a more accurate prediction for the void distribution function over a large range of scales (see Figure~\ref{Fig:VSF}). This brings the application of the distribution function of voids for precise cosmological measurements a step closer.

The model is derived for the matter density field. When using tracers of dark matter, such as dark matter halos or galaxies, to define voids, as is the case in real observations, \citet{Ronconi2019, Contarini2019} show that the model still works well, provided that the density contrast between the tracers versus dark matter is calibrated. 

\subsection{Applications}

The VSF has been proposed as a means to test theories of gravity. For modified theories of gravity, in the standard excursion set theory, $\delta_{v}$ can depend on scale and environmental density \citep{Clampitt2013, Voivodic2017}. This leads to different VSFs compared to the VSF in general relativity. In coupled scalar field models for example, due to the `screening mechanism', the strength of gravity may vary depending on the density environment. The dynamics of voids are particularly sensitive to it. This may lead to different abundances of voids in those models \citep{Clampitt2013, Lam2015, Voivodic2017}. Predictions for the VSFs from the $f(R)$ and symmetron models from $N$-body simulations by \cite{Cai2015, Voivodic2017} qualitatively confirm this. However, it was also shown that when mock galaxies are used to define voids, it becomes challenging to distinguish different models \citep{Voivodic2017}.

The sum of neutrino masses has been shown to leave its signature in the VSF. Using cosmological $N$-body simulations, \citet{Massara2015} shows that voids in massive neutrino cosmologies are less evolved than those without neutrinos: there is a larger number of small voids and a smaller number of large ones. \citet{Kreisch2019} shows that increasing the mass of the neutrinos increases the number of small dark matter voids and decreases the abundance of large voids. The opposite effect is found when using highly biased dark matter halos to define voids. The halo mass function in void environments is also found to be more sensitive to the sum of the masses of neutrinos \citep{Zhang2020}. Combining the VSF with the halo mass function and power spectrum can also help to tighten the constraints on the mass of neutrinos \citep{Bayer2021}. In observations, analyzes with data from the SDSS-BOSS galaxy sample suggests that having void statistics adds modest but nonzero information beyond galaxy clustering. Void statistics may also be effective at constraining the sum of neutrino masses from below \citep{Thiele2024}. The VSF at high redshifts could also be useful to break the degeneracy between the sum of neutrino masses and the strength of gravity in the $f(R)$ model \citep{Contarini2021}.

In observations, \citet{Sahlen2016} used extreme-value statistics on the largest void from the 2MASS-WISE galaxy catalogue and cluster from the Atacama Cosmology Telescope (ACT) to constrain $\Omega_m$ and $\sigma_8$, the matter density parameter and the density dispersion of the linear-theory density field in 8 Mpc$/h$ spheres. \citet{Contarini2023,Contarini2024} used voids from the SDSS-BOSS-DR12 sample to obtain $\sim 10\%$ constraints in  $\Omega_m$, $\sigma_8$, and the parameter for the dark energy equation of state. They have also demonstrated that compared to the halo mass function, the VSF is sensitive to density perturbations on larger scales, they are therefore complementary to each other \citep{Pelliciari2023}. A similar level of precision for cosmological constraints was obtained using voids in the SDSS-DR7 sample \citep{Fernandez-Garcia2024}, fitted to the model developed in \citet{Betancort-Rijo2009}.

\citet{Sahlen2016b} demonstrated that combining void abundance with cluster abundance can break degeneracies among cosmological parameters and offers constraints on modified gravity via the linear growth rate index $\gamma$ from $f=\Omega_m^{\gamma}$. \citet{Sahlen2019} further showed that the combination of VSF and halo mass function is able to break the degeneracy between the dark energy equation of state and the sum of neutrino masses. Using a simulation-based modelling framework, \citet{Thiele2024} has shown that the VSF adds modest extra constraints to the sum of neutrino mass. The constraint for the dark energy equation of state with Euclid has been predicted to be below 10\% \citep{Contarini2022}. Percent-level constraints for cosmological parameters with the VSF from the CSST survey are expected \citep{Song2024}.
\begin{figure*}
\hspace{-0.4cm}
\centering
\includegraphics[width=1.0\columnwidth]{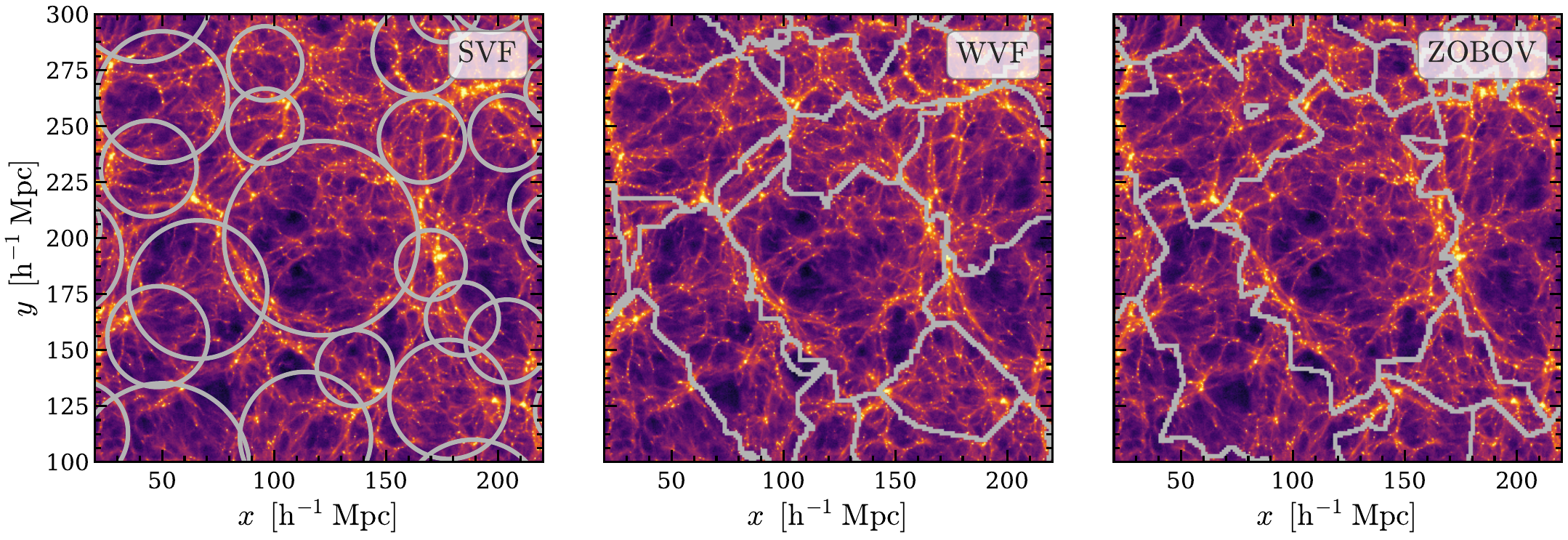}
\caption{Comparing three different void finding algorithms for voids identified in the same cosmological $N$-body simulation. The spherical void finder (SVF) \citep{Padilla2005}, the watershed void finder (WVF) \citep{Platen2007} and the ZOBOV void finder \citep{zobov} are shown on the left, the middle and the right. Grey curves representing boundaries of voids are drawn on top of the projected density of dark matter. Despite some detailed differences for the exact borders of voids, the largest void regions in the simulation are consistently found by all the three algorithms. This figure is taken from Figure~2 of \citet{Paillas2019}.}
\label{fig:void_2D}
\end{figure*}

\subsection{Finding voids in simulations and observations}

In simulations and observations, individual voids have a wide variety of shapes, depths and density profiles. There is no clear definition for what should be called a void. It is unambiguous that they should be under-dense, but how to define their centres, where are their boundaries and how underdense should they be, are all somewhat uncertain. The flexibility of their definition has led to the development of a range of algorithms to find them. Figure~\ref{fig:void_2D} compares voids identified in the same simulation using three different void finding algorithms, the spherical void finder \citep{Padilla2005}, the watershed void finder \citep{Platen2007} and ZOBOV \citep{zobov}. We can see that the largest voids are found by all the three algorithms, but their boundaries and centres are all different. In addition, small voids found in one algorithm ({\sc SVF}) may not be identified in another algorithm (ZOBOV). These differences indicate that the statistical properties of voids may depend on the void finding algorithm. The cosmological applications of them may need to account for this.

There are two main categories of void finding algorithms: underdensity-finding, and dynamical. The first uses the distribution of matter or tracers of matter to define under-dense regions, such as the spherical void finding algorithm \citep[e.g.][]{Hoyle2002,Padilla2005}, watershed based algorithms \citep{Colberg2005, Platen2007, zobov} and the {\sc Popcorn} void finding algorithm \citep{Paz2023}. The spherical void finder identifies regions of the Universe which satisfy a void density threshold, such as the one derived from the spherical evolution model $\delta_v=-0.8$. ZOBOV locates density depressions using a Voronoi tessellation to measure density at the location of each galaxy. From the lowest density depressions, a watershed algorithm is applied to group neighbouring Voronoi cells into `zones' to form voids \citep{zobov}.

The second category uses dynamical quantities such as the velocity divergence and phenomena such as stream-crossing to identify voids. For example, with the Hessian matrix of the smoothed density field, or the gravitational potential field, one can identify voids as regions with all three eigenvalues below a certain threshold \citep[e.g.][]{Hahn2007, Forero-Romero2009, lavauxwandelt2010,Aragon-Calvo2010, Hoffman2012, Cautun2014}. The scale of smoothing and the thresholds for the eigenvalues are free parameters in these algorithms. Some dynamical methods are arguably physically objective, but rely on information available only in simulations, not straightforwardly in observations. \citet{Shandarin2012} combined information of density and velocity i.e., the full 6D phase space information to classify the cosmic web, including voids. \citet{Falck2012} classified particles in simulations as `void' particles if they undergo no crossings with respect to any other particles over the course of a simulation. In some cases, the void particles may not be easily separable into idealized, nearly convex or polyhedral shapes, but instead may be better thought of a void region that percolates wherever galaxies, filaments and walls are not \citep{Falck2015,StueckerEtal2018}

In general, the characteristic of voids being under-dense converges among different algorithms, but the quantitative measurement for the abundance of voids as a function of radius may differ significantly. Which algorithm to choose may depend on the specific application for cosmology or astrophysics. A summary comparing different algorithms can be found in \citet{Colberg2008} and also in \citet{Libeskind2018} in the context of the cosmic web.

In observations, samples of voids have been identified in major galaxy redshift surveys and lensing surveys. These include voids in the CfA survey \citep{Vogeley1991}, the Point Source Catalog redshift (PSCz) survey \citep{Hoyle2002},  the Two-Degree Field Galaxy Redshift Survey \citep{Hoyle2004}, and the Dark Energy Survey (DES) \citep{Sanchez2017, Kovacs2022DESY3_VoidLensing}. Perhaps the most exploited dataset for voids is in the SDSS area \citep{Pan2012, Sutter2012, Cai2014, Nadathur2016, Mao2017, Zhao2020, Zhao2022, Douglass2023, Tamone2023}. These have been used to constrain cosmological parameters via redshift-space distortions \citep{Kaiser1987} and the Alcock-Paczy\'nski (AP) effect \citep{alcockpaczynski1979}, 
measure the growth of structure, constrain the sum of neutrino masses via the effect of gravitational lensing, and study the imprints of voids on the CMB via the integrated Sachs-Wolfe (ISW) effect \citep{Sachs1967}.

\begin{figure}
\vspace{-3.5 cm}
\begin{centering}
\includegraphics[width=0.48\columnwidth]{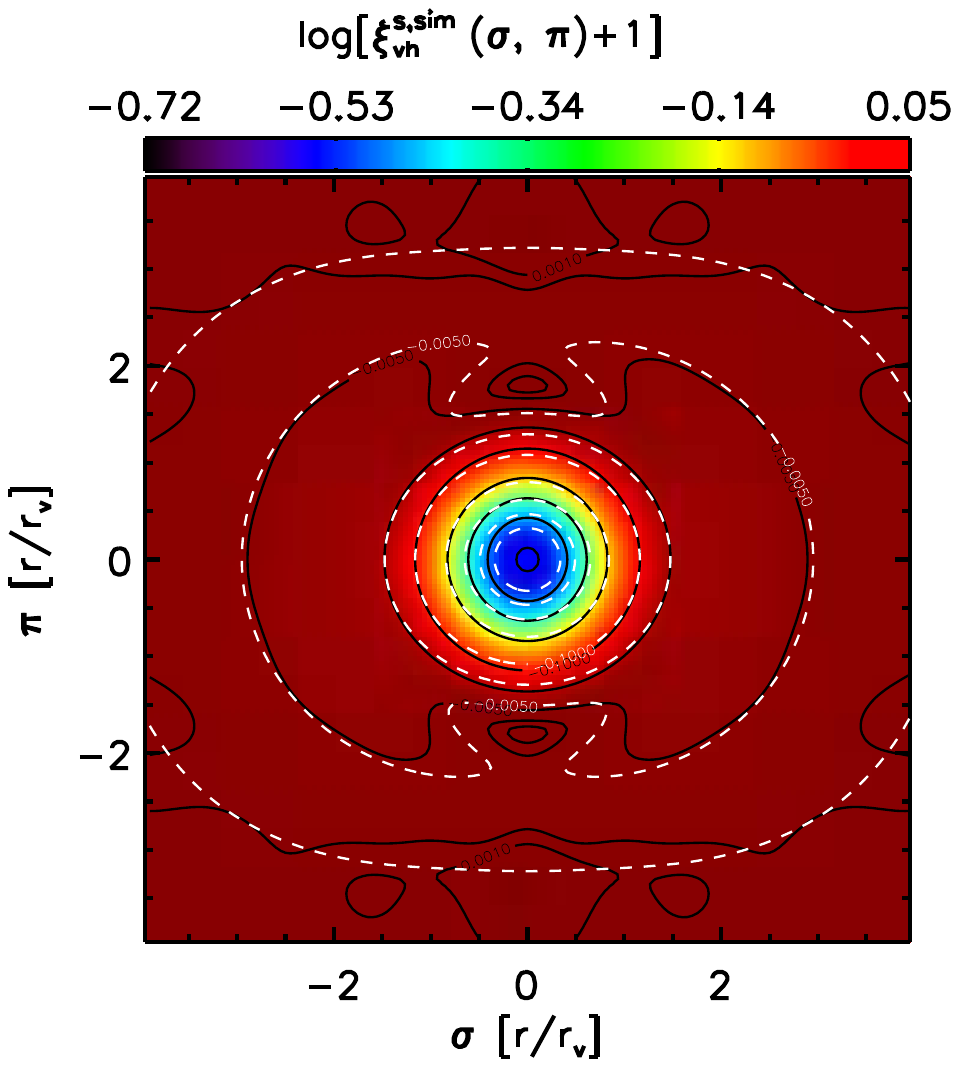}
\includegraphics[width=0.48\columnwidth]{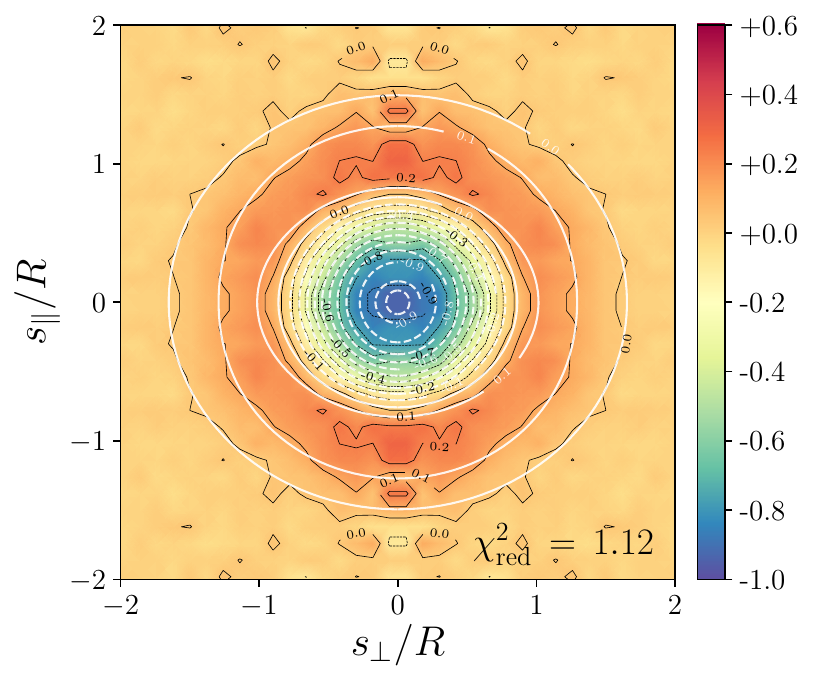}
\includegraphics[width=0.48\columnwidth]{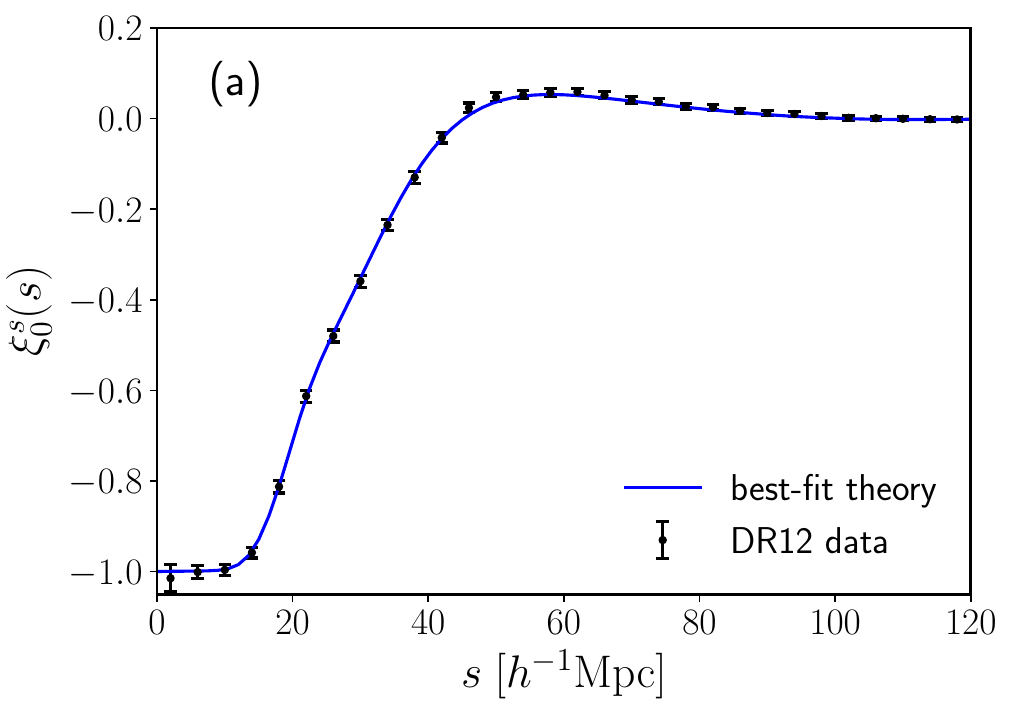}
\includegraphics[width=0.48\columnwidth]{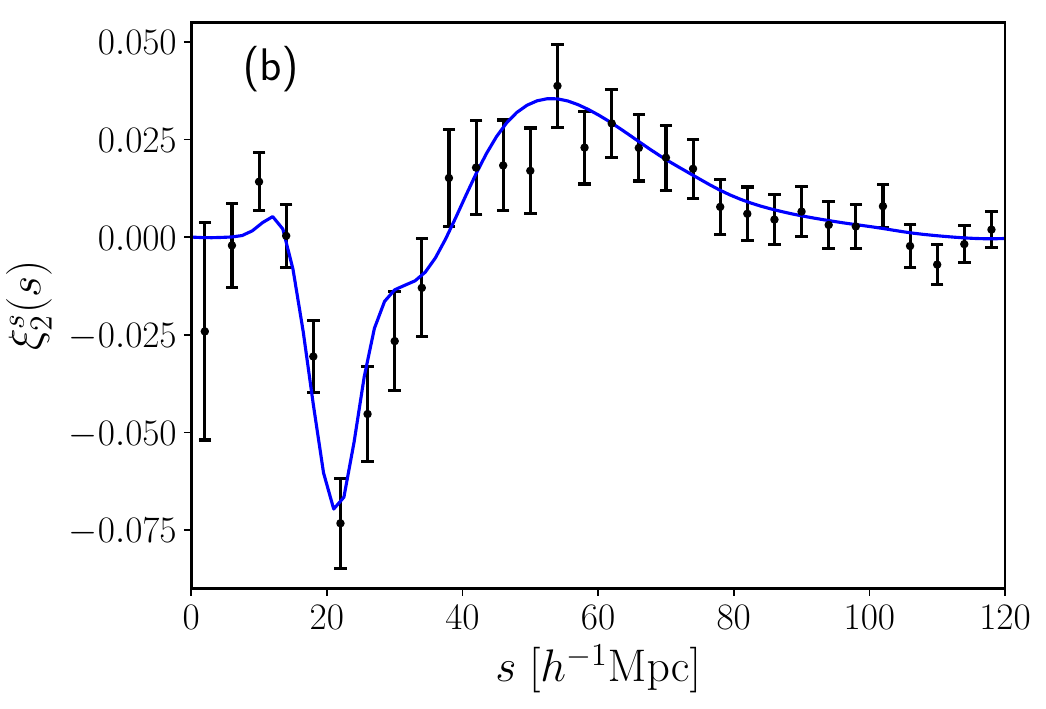}
\end{centering}
\caption{Top-left: comparing the void-halo correlation function in redshift space measured from simulations with linear theory. The black contours are results from the simulation, and the white contours are the best-fit linear model \citep{Cai2016}.
Top-right: void-galaxy correlation function measured from the SDSS-BOSS-DR12 galaxy sample. The black contours are results from observation, and the white contours are the best-fit model from linear theory. The inverse Abel transformation was used for reconstructing the real-space void-galaxy correlation function \citep{Hamaus2020}. Bottom-left and bottom-right: the monopole and quadrupole moments of the void-galaxy correlation function measured with the SDSS-BOSS-DR12-CMASS galaxy sample. Data points with errors are from observations; the blue lines are the best-fit model based on the Gaussian streaming model. Reconstruction with the Zeldovich approximation was used to recover the real-space void-galaxy correlation function \citep{Nadathur2020}. From fitting to the measurements, a few percent-level constraints for the growth parameter and the AP parameters have been achieved \citep{Hamaus2020,Nadathur2020}. These plots are taken from the above three papers mentioned.}
\label{fig:void_RSD_theory}
\end{figure}

\section{Redshift-space distortions around voids}
\label{sec: RSD}

Galaxy redshift surveys such as SDSS and DESI \citep{DESI2016} provide the 3D distribution of millions of galaxies over a large volume of the Universe. Voids can be identified in these surveys from finding under densities in the galaxy number density field, which is thought to be tracing the underlying large-scale structure of matter. Stacking the galaxy number densities around void centres yields the void-galaxy correlation function $\xi_{vg}(s_{\perp}, s_{\parallel})$, where $s_{\perp}$ and $s_{\parallel}$ are the redshift-space transverse and line-of-sight distances from the centre of the voids. When averaging over a large sample of voids, the real-space void-galaxy correlation function $\xi_{vg}(r)$ should be isotropic, but $\xi_{vg}(s_{\perp}, s_{\parallel})$ is anisotropic due to (i) redshift-space distortions \citep{Kaiser1987}, and (ii) the Alcock-Paczynski (AP) effect \citep{alcockpaczynski1979}. The joint analyzes of these two effects with the observed void-galaxy correlation function provides complementary constraints for cosmological parameters. It has emerged as a major cosmological probe. 

\subsection{Dynamical distortions}
The effect of redshift-space distortions arises from the fact that the observed line-of-sight distance of a galaxy $s_{\parallel}$ is perturbed by its peculiar velocity $v^{\rm pec}_{\parallel}$ along the line of sight
\begin{equation}
s_{\parallel}=r_{\parallel}+v^{\rm pec}_{\parallel}/[aH(a)], 
\end{equation}
where $r_{\parallel}$ is the real-space comoving distances of a galaxy, $a$ is the scale factor of the Universe and $H(a)$ is the Hubble parameter at $a$. This effect leads to dynamical distortions for the observed void-galaxy correlation function. The distortion is small, but is detectable with a large sample of voids. 

The mapping between the redshift-space void-galaxy correlation function $\xi_{vg}(s_{\perp}, s_{\parallel})$ and its real-space counterpart $\xi_{vg}(r)$ encodes information about peculiar velocities.
To extract cosmological information from this measurement, we need a model to connect them. The number of galaxies around voids is conserved in the mapping from real to redshift space. With the plane-parallel approximation where $v^{\rm pec}_{\parallel}/[aH(a)] \ll 1$, keeping all terms to the linear order, $\xi_{vg}(r)\ll 1 $ and $\partial v(r)/\partial r \ll aH(a)$ \citep{Cai2016},
\begin{equation}
\label{Eq:zspaceProfile}
\xi_{vg}(s_{\perp}, s_{\parallel})=\xi^r(r)+\frac{1}{3}\beta\bar{\xi}^r(r)+\beta\mu^2[\xi^r(r)-\bar{\xi}^r(r)], 
\end{equation} 
where $\mu=\cos\theta$ and $\theta$ is the subtended angle from the line of sight; $\beta=f/b$ and $f=d\ln D/d\ln a$ is the linear growth rate with $D$ being the linear growth factor and $b$ is the linear galaxy bias, a factor of proportionality between the amplitude of clustering in galaxies and matter. $\bar\xi^r(r)=\frac{3}{r^3}\int_0^r\xi^r(r')r'^2dr'$ is the volume averaged real space void-galaxy correlation function. Note that $\xi_{vg}(r)$ is the real-space density profile of voids traced out by galaxies. In linear theory, $\xi_{vg}(r)=b\delta(r)$, and so $\bar\xi^r(r)$ is coupled to the peculiar velocity via Equation~\ref{eq:LinearVelocity}. An example of the model and its comparison with measurements from simulations and observations is shown in the top panels of Figure~\ref{fig:void_RSD_theory}. The model thus encodes information about the peculiar velocity, which is related to the growth parameter $\beta$. This enables us to constrain the growth parameter.  

Linear theory applies to regions where the density fluctuation is small. This is a good approximation on large scales, but it may not be the case near the centre of voids. For the more general cases, the streaming model \citep{Peebles1980} is exact:
\begin{equation}
\label{Eq:StreamingModel}
1+\xi_{vg}(s_{\perp}, s_{\parallel})=\int [1+\xi^r(r)]p(r,v^{pec}_{\parallel})dv_{\parallel}^{pec}, 
\end{equation} 
but it requires knowing the real-space correlation function and the full velocity distribution function $p(r,v^{pec}_{\parallel})$ \citep{Peebles1980, Paillas2021}, which is challenging. By assuming that the velocity distribution function is Gaussian, the above equation becomes
\begin{equation}
\label{Eq:StreamingModel}
1+\xi_{vg}(s_{\perp}, s_{\parallel})=\int [1+\xi^r(r)] \frac{1}{\sqrt{2\pi\sigma^2_v(r,\mu)}} \exp{\left[-\frac{v_{\parallel}-v^{pec}(r)\mu }{2\sigma^2_v(r, \mu)}\right] dv_{\parallel}^{pec}}, 
\end{equation} 
where $v^{pec}(r)$ is the radial peculiar velocity profile, and $\sigma^2_v(r, \mu)$ is the velocity dispersion. This is the so-called Gaussian streaming model \citep{Fisher1995}. It was shown to be more accurate than the linear one down to relatively small scales \citep{Paz2013,Hamaus2015, Cai2016, Nadathur2016, Nadathur2017, Paillas2021}. 

\subsection{Geometric distortions}
In addition to the dynamical distortion, geometric distortions around voids arise when converting the observed redshifts of galaxies into distances \citep{lavauxwandelt2012, Hamaus2014}. To do so, we need to assume a fiducial cosmological model to have the distance-redshift relationship. If the fiducial model is different from the true underlying cosmology of the Universe, all the distances will be stretched, an effect known as Alcock-Paczynski (AP) distortions \citep{alcockpaczynski1979}.
The AP effect amounts to a universal re-scaling for the line-of-sight and transverse distances compared to the fiducial cosmology \citep{Ballinger1996}: 
$s_{\perp}=q_{\perp}s'_{\perp}$ \& $s_{\parallel}=q_{\parallel}s'_{\parallel}$, where the prime represent quantities in the fiducial cosmology. The two scaling factors are related to cosmological parameters via 
$q_{\perp}=D_M/D_M'$ and $q_{\parallel}=H/H'$ where $D_M$ is the angular diameter distance and $H$ is the Hubble parameter.

The AP effect enters the observed void-galaxy correlation function (e.g. Equation~\ref{Eq:zspaceProfile}) through coordinate transformations: $\xi^s(s_{\perp}, s_{\parallel})=\xi^s(q_{\perp}s'_{\perp}, q_{\parallel}s'_{\parallel})$. It interferes with the interpretations of the dynamical distortion, but the two effects are not degenerate. In cosmological parameter inferences, both effects will be modelled in $\xi^s(s_{\perp}, s_{\parallel})$. This enables us to constrain cosmological parameters by comparing the model against observations. Examples of this kind of analysis are shown in Figure~\ref{fig:void_RSD_theory}.

\subsection{Applications}
Redshift-space distortions around voids encode information about the density and velocity. This is ideal for capturing the difference between general relativity and non-standard gravity models, where violation of the equivalence principle may occur most evidently in voids. Therefore, redshift-space distortions around voids is a promising observable which can capture the combination of these two effects \citep{Cai2015, Cai2016}. Recent tests with mock catalogues confirm the physical scenario, but find that at the statistical precision offered by a volume of $\sim 1$ (Gpc/$h$)$^3$, the linear growth parameter fitted from redshift-space distortions can not distinguish between general relativity and their specific choice of the $f(R)$ and nDGP models \citep{Wilson2023}. \citet{Mauland2023} used the void-galaxy correlation function to investigate the degeneracy between massive neutrinos and $f(R)$ modified gravity, but concluded that at the accuracy of a few percent, the present modelling is insufficiently accurate to do so. 

In observations, analyzes of redshift-space distortions around voids to constrain the growth rate parameter have been reported in \citet{Achitouv2017b} from the 6dF Galaxy Survey, and \citet{Hawken2017} from the VIPERS survey, delivering constraints for the growth parameter on the order of $\sim 10\%$. The joint analysis of redshift-space distortions and the AP distortions has emerged to become a major cosmological probe. It has been applied to data from the SDSS-BOSS and eBOSS surveys from multiple research groups with different methods \citep{Hamaus2016,Hamaus2017, Achitouv2019, Nadathur2019, Hamaus2020,Hawken2020,Nadathur2020,Aubert2022,Woodfinden2022, Woodfinden2023, Fraser2024}. Despite the differences in the choice of modelling among different research teams, most of them are able to bring down the constraints on the growth parameter to a few percents level (see Figure~\ref{fig:void_RSD_theory} for results from different analyzes of a similar dataset.). This is complementary and comparable to the standard two-point statistics. This may indicate that like other conventional analyzes of large-scale structure, the RSD and AP analyzes around voids are already limited by systematic errors, including observational systematics and limitations in the accuracy of our modelling. 

A common underlying assumption for all the above models of redshift-space distortions is that the centres of voids do not move. This is not exactly the case. Void centres do have peculiar velocities, and this complicates the modelling of redshift-space distortions \citep{Chuang2017, Nadathur2019b}. In addition, \citet{Correa2022} points out the void-galaxy correlation may also be complicated by the ellipticity of voids. Two approaches have been developed to deal with this, by reconstructing the real-space void-galaxy correlation function from data itself. \citet{Hamaus2020} deproject the observed projected correlation void-galaxy correlation function; \citet{Nadathur2019b} proposes using velocity reconstructions to remove redshift-space distortions. Both have been demonstrated to deliver unbiased constraints for cosmological parameters with mock catalogues, and have set some of the tightest constraints for the growth parameters with data from BOSS \citep{Hamaus2020, Nadathur2020}. They have also been predicted to deliver percent-level constraints for the growth parameter and sub-percent-level constraints for the AP parameters with Euclid \citep{Hamaus2022,Radinovic2023}. 

It is also worth mentioning that the void-void clustering, together with the void-galaxy clustering has been shown to provide extra information about the baryon acoustic oscillations (BAO), these have helped to tighten the constraints for the AP parameters  \citep{Kitaura2016, Liang2016, Zhao2020, Zhao2022, Tamone2023}.

\begin{figure}
\centering
\resizebox{100mm}{!}{
\includegraphics[angle=270]{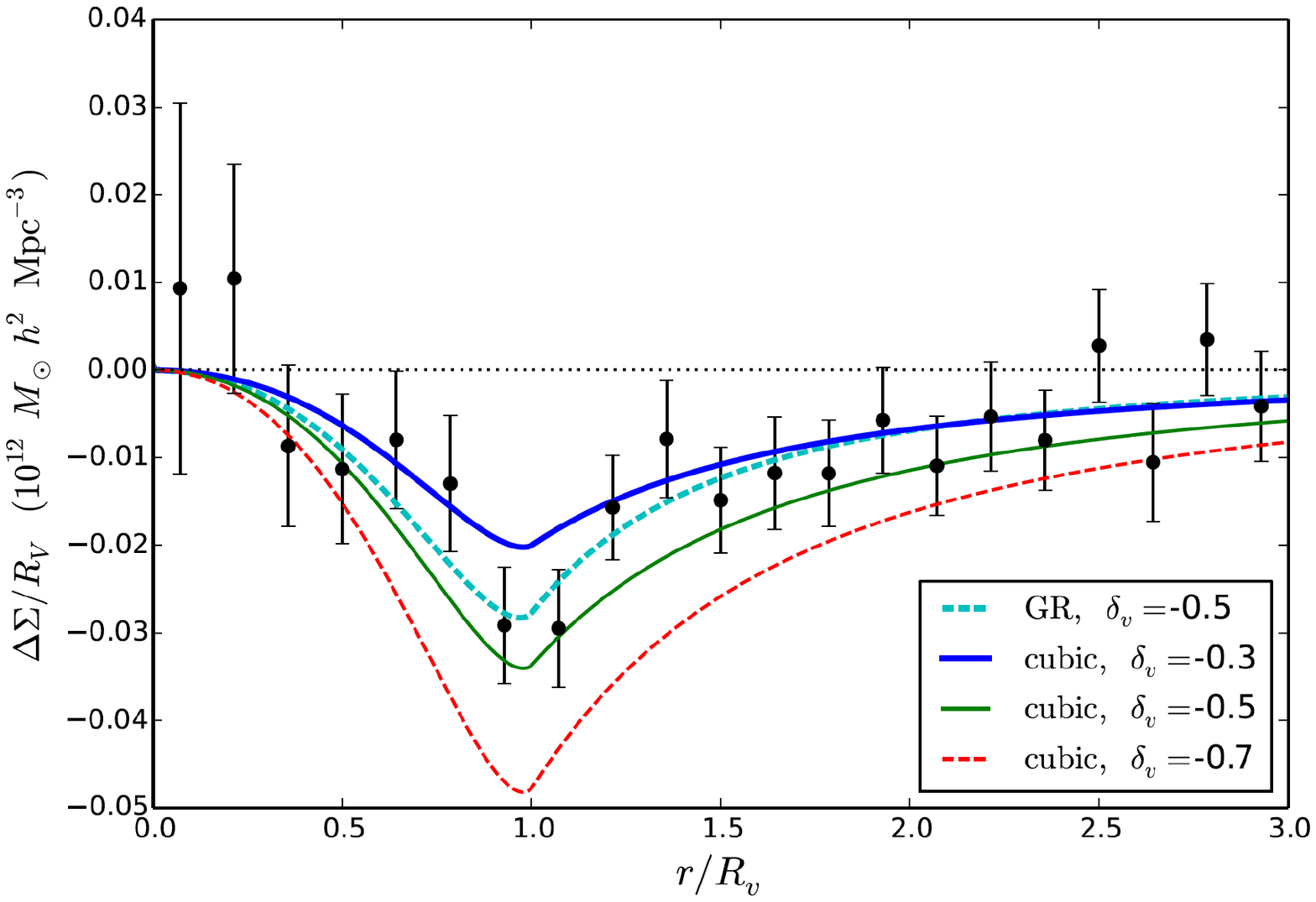}}
\caption{Comparing the observed lensing tangential shear profiles (data points with errors) around SDSS voids using the SDSS LRG galaxy image data with different models. Coloured lines are model predictions from general relativity and the Cubic Galileon model with different central densities of voids, as indicated by the legend. The profiles are normalised by the radius of each void, as indicated in the x-axis. The figure is taken from Figure~3 of \citep{Baker2018}.}
\label{fig:void_lensing_MG}
\end{figure}

\begin{figure}
\vspace{-1.5 cm}
\begin{centering}
\includegraphics[width=0.7\columnwidth, angle=270]{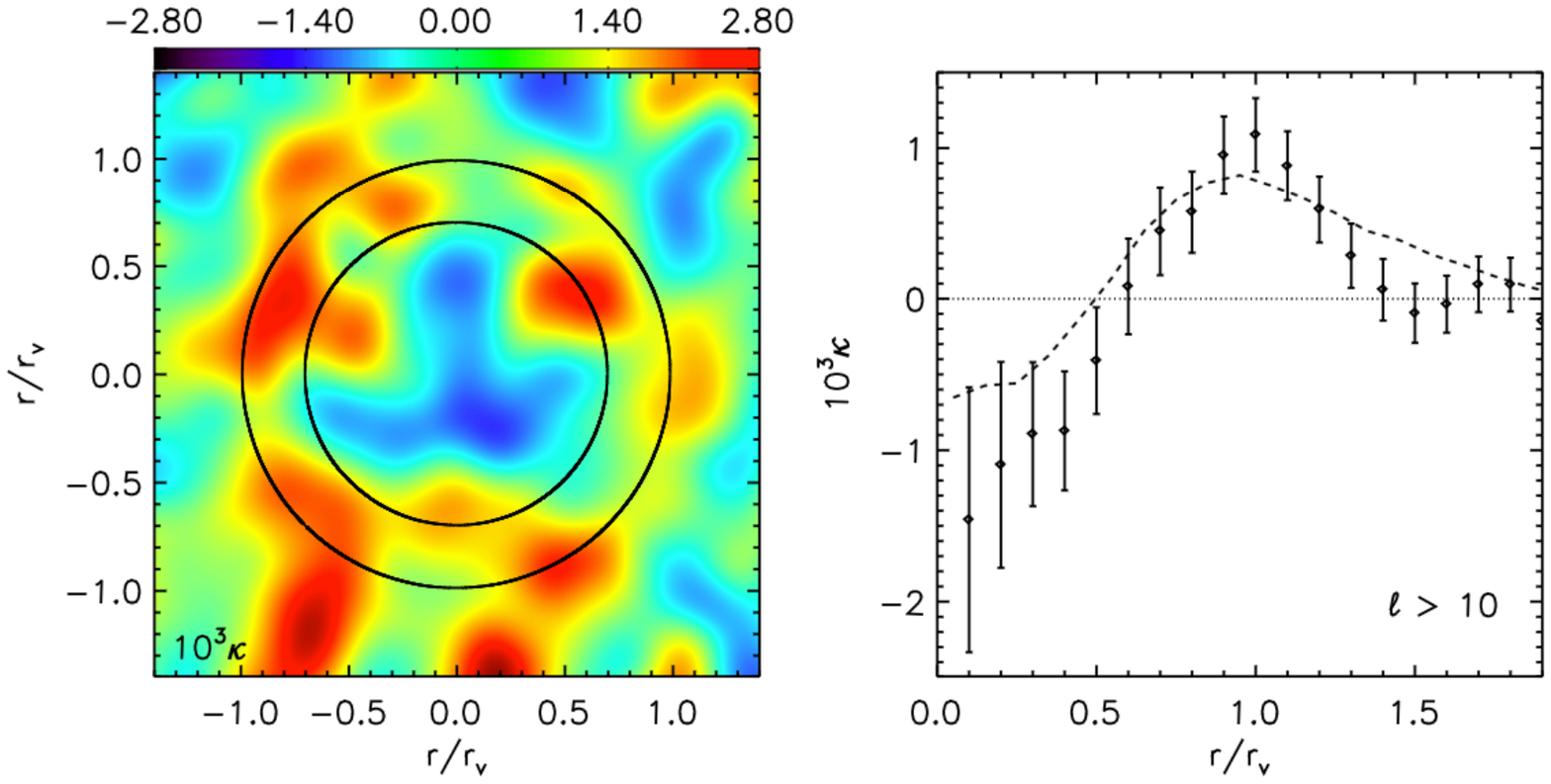}
\end{centering}
\vspace{-1 cm}
\caption{Left: the stacked map of CMB lensing convergence $\kappa$ from Planck \citep{Plancklensing2014} with the locations of voids found in the SDSS-BOSS-DR12-CMASS galaxy sample. Voids with the radius $r_v > $20 Mpc/$h$ are used for the stacking. The inner circle of radius and equal-area outer ring mark the extent of the compensated filter, normalised with $r_v$. Right: the averaged convergences from the centre of the stacking on the left. Data points with errors are measurements from observations; the dashed line is the measurement from mock catalogues from the standard $\Lambda$CDM model, which agrees with the observational data within the errors \citep{Cai2017}. The figure is taken from Figure~3 of \citet{Cai2017}. }
\label{fig:void_lensing_CMB}
\end{figure}

\section{Gravitational lensing by voids}
\label{sec: lensing}

Cosmic voids can be substantial gravitational potential hills, which generally deflect photons passing through them. The spatial and temporal variations of the metric gravitational potentials $\Phi(x,t)+\Psi(x,t)$ generates observational effects on photons. The spatial gradients in $\Phi+\Psi$  give rise to the gravitational lensing effect around voids, which is characterised by the lensing potential:
\begin{equation}
\psi_{\rm lens}=\frac{1}{c^2}\int \frac{D_{ls}}{D_sD_l} (\Phi+\Psi) dz, 
\end{equation}
with $\Phi$ and $\Psi$ being respectively the time and space parts of the metric potentials in the perturbed FRLW metric. $\psi_{\rm lens}$ is defined as the lensing potential, $z$ is the redshift, and $D_l$, $D_s$ and $D_{ls}$ are the line-of-sight angular diameter distances of the lens, the source and that between the lens and the source, respectively. 

The 2D Laplacian of the lensing potential is related to the lensing convergence $\kappa$ via $\kappa = \frac{1}{2} \nabla^2 \psi_{\rm lens}$. In general relativity where $\Phi=\Psi$, it can be expressed as
\begin{equation}
\kappa=\frac{3H_0^2\Omega_m}{2c^2} \int  \frac{D_l D_{ls}}{D_s}\frac{\delta}{a} dz.
\end{equation}
The lensing convergence is therefore related to the projected mass density contrast $\delta$ along the line of sight. Gravitational lensing by voids therefore provides a means to measure the projected mass distribution around voids.

The gravitational lensing by voids affects any light shining from behind the voids, from the cosmic microwave background and galaxies. The effect can be detected by observing the coherent distortions of the CMB temperature map, called CMB lensing; or the coherent shear signal on the images of background galaxies, which may be called void-galaxy lensing.

Cross-correlations between voids with a reconstructed map of the lensing convergence, either coming from the CMB, or from the surveys of galaxy images, offers a direct way to measure the projected mass density around voids. Alternatively, the projected density profiles of voids can be measured by directly stacking the shear signal of the background galaxies around them. The lensing convergence profile is related to the tangential component of the shear $\gamma_t$ via,
\begin{equation}
\gamma_t(R_p)=\kappa(<R_p)-\kappa(R_p),
\end{equation}
where $\kappa(R_p)$ and $\kappa(<R_p)$ are the average convergence at the projected radius $R_p$ and within $R_p$ from the void centre. Figure~\ref{fig:void_lensing_MG} shows an example of the lensing tangential shear profiles around voids measured from the SDSS LRG galaxy sample. Figure~\ref{fig:void_lensing_CMB} shows the first detection of the CMB lensing convergence around voids from the SDSS-CMASS sample.

A prediction for lensing by individual voids was made in \citet{Amendola1999}, concluding that only voids with radius greater than 100Mpc/$h$ will have the signal-to-noise ratio of unity. The first forecasts for the detectability of void lensing via stacking was made by Krause et al. \citep{Krause2013} and Higuchi et al. \citep{Higuchi2013}. Detections of the signal from lensing surveys were first made in the SDSS area \citep{melchioretal2014, clampittjain2014}, and later with the DES science verification data \citep{Gruen2016,Sanchez2017} and year-1 data \citep{Fang2019,Vielzeuf2021}, and at high redshifts with the Lyman$-\alpha$ forest tomography in the COSMOS field \citep{Krolewski2018}, and in more recent data releases of the DES survey. 

CMB lensing of voids has been detected by stacking the reconstructed lensing convergence from CMB lensing, first in the SDSS area \citep{Cai2017,Chantavat2017, Raghunathan2020}. Detections have been made using the cross-correlations of other major galaxy surveys: the WISE-Pan-STARSS data set \citep{Camacho-Ciurana2023}, the DES area \citep{Kovacs2022DESY3_VoidLensing, Demirbozan2024} and the DESI legacy survey data \citep{Hang2021,Dong2021}. The signal-to-noise ratio of lensing by voids is now sufficiently significant that it has been used to constrain the amplitude parameter of lensing convergence $A_{\kappa}$, as a consistency check between the linear growth parameter and the effect of gravitational lensing. 

\subsection{Applications}
Gravitational lensing by voids is a major way to measure the total matter density profile of voids. Despite the line-of-sight projection effect, which erases some information from the full 3D voids, most cosmological applications of individual voids mentioned in Section~\ref{sec:Individual_Voids} rely on the measurements of lensing. The effect is therefore expected to be powerful for constraining the sum of neutrino masses and testing theories of gravity.

Neutrinos have a higher mass fraction in voids, since they are diffuse compared to other matter. So, the impact of neutrinos on the growth of structure is expected to be stronger in voids than in high-density regions. Forecasts with simulations have shown that the profiles of voids are indeed a sensitive probe of neutrino masses \citep[e.g.][]{Massara2015, Coulton2020}, and that gravitational lensing by voids has the potential to set constraints on the sum of neutrino masses \citep[e.g.][]{Vielzeuf2023}. Forecasts for the Euclid survey also indicate that lensing of voids can provide better constraints on other cosmological parameters when combined with other cosmological probes \citep{Bonici2023}.

For theories of gravity with coupling between the strength of gravity and matter density, the density profiles of voids will be altered compared to their version in standard gravity. This can be detected through lensing of voids \citep{Clampitt2013, Cai2015, Cautun2018, Davies2019}. For others, such as the Galileon models (or the massive gravity model), the relation between the lensing potential and the density perturbation can be modified. The lensing signal around voids can be very different, even with the same mass density as in general relativity. Lensing of voids can therefore be used to distinguish them from general relativity \citep{Barreira2015,Baker2018, Su2023}. An example of this kind is given in Figure~\ref{fig:void_lensing_MG}. When combined with the two-point correlation function, lensing of voids is also useful to break the  degeneracy among cosmological parameters \citep{Davies2021}. In addition, the effect of gravitational lensing probes the sum of the two metric potentials, the motions of galaxies respond only to the Newtonian potential. The combination of lensing and redshift-space distortions can be used to test the gravitational slip \citep[e.g.][]{Zhang2007,Reyes2010,Alam2017,Amon2018,Singh2018,Skara2020, Wenzl2024}. This applies to voids, where the gravitational slip, if any, may be stronger \citep{Spolyar2013}. 

An extended version of lensing around voids is `troughs', which measures the lensing signal around under densities in 2D projected galaxy number density fields \citep{Gruen2016, Barreira2017, Friedrich2018, Gruen2018, Brouwer2018}. An analytical model to predict the signal has been developed in \citet{Gruen2016, Friedrich2018}, which allows for constraints on cosmological parameters. The method has been applied to data from the Dark Energy Survey \citep{Gruen2016,Gruen2018}, and the Kilo-Degree Survey \citep{Burger2023}, delivering complimentary and comparable constraints on $\Omega_m$-$\sigma_8$ and the galaxy bias parameter $b$ compared to the conventional lensing 3$\times$2 analyzes of the same data.

\section{The imprints of voids on the cosmic microwave background}
The spatial perturbation of the gravitational potentials by voids causes gravitational lensing effects of the CMB, as discussed in the previous section. 
The temporal perturbation of the potentials induced by voids alters the energy of photons traversing them. This induces additional temperature fluctuations of the CMB, known as the Integrated Sachs-Wolfe (ISW) effect \citep{Sachs1967}:
\begin{equation}
\label{eq:ISW}
\frac{\Delta T}{T_{\rm CMB}} = -\frac{1}{c^2}\int (\dot \Phi+ \dot \Psi) dt,
\end{equation}
where the $\dot \Phi$ and $\dot \Psi$ are the time derivative of the metric potentials and $T_{\rm CMB}$ is the CMB temperature. In a $\Lambda$CDM universe, the accelerating expansion of the Universe caused by dark energy stretches cosmic voids and superclusters, causing their gravitational potentials to decay. CMB photons travelling through them will lose/gain energy. The observed CMB temperature along the line of sight of a void/supercluster will decrease/increase by the amount specified by Equation~\ref{eq:ISW}. A detection of the ISW signal can therefore provide constraints on dark energy. It can also be used to constrain theories of gravity \citep{Song2007, Lombriser2012,Barreira2014, Renk2017}.

\begin{figure}
\begin{centering}
\includegraphics[width=1.0\columnwidth]{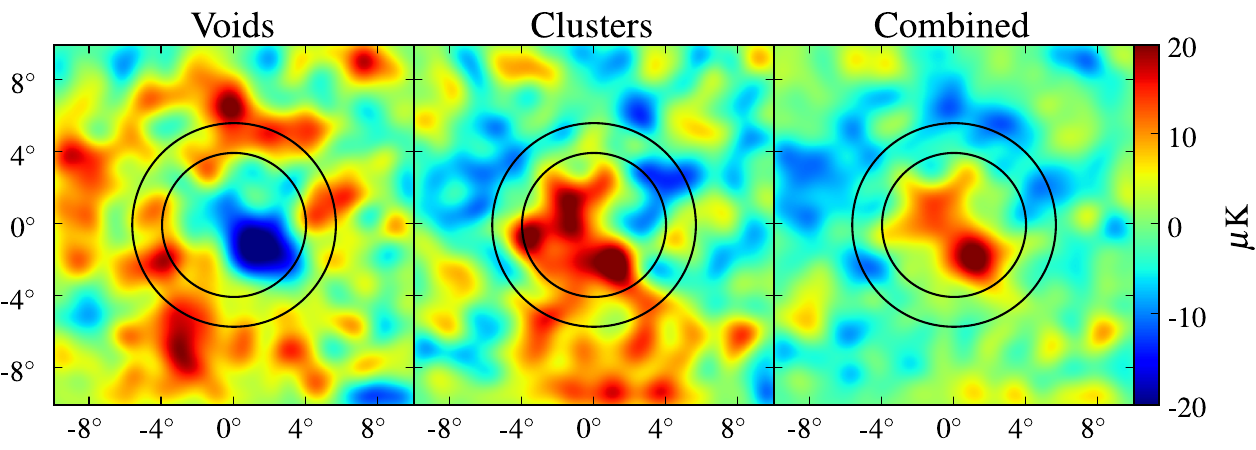}
\end{centering}
\caption{Stacked regions on the temperature map of the CMB corresponding to 50 voids and 50 superclusters identified in the SDSS Luminous Red Galaxy (LRG) sample. Cold and hot spots appear in the void and cluster stacks, respectively, with a characteristic radius of four degrees, corresponding to the scale of 100 Mpc/$h$. The sign of the signal from voids is flipped and averaged with the signal from the superclusters, shown as `Combined'. These are expected if dark energy is causing the gravitational potentials of voids and superclusters to decay -- the observation can be taken as a piece of evidence of dark energy. However, the amplitude of the signal, which is at the order of 10 $\mu K$, is a few times higher than expected in the standard $\Lambda$CDM model \citep{Granett2008}. The figure is taken from Figure~1 of \citet{Granett2008}.}
\label{Fig:ISW}
\end{figure}

In the standard $\Lambda$CDM model, the ISW signal is much smaller than the primordial CMB temperature fluctuation. This is a challenge for its direct detection. One way to detect the signal is by stacking the CMB temperature map with voids and superclusters found in galaxy redshift surveys. This should yield a cold and hot spot respectively for voids and superclusters. It was first conducted in \citet{Granett2008} where a somewhat unexpected high significance ($\sim4\sigma$) result was reported by stacking superstructures found from the photo-$z$ galaxy catalogue in the SDSS area (see Figure~\ref{Fig:ISW} for the main results). However, the amplitudes of the signal were higher than expected from the standard $\Lambda$CDM model. Follow-up analysis using similar methods with spectroscopic redshift catalogues (but still in the SDSS area) confirms the results qualitatively \citep{nadathuretal2012,flenderetal2013, Ilic2014, Cai2014, Kovacs2015, PlanckISW2016, aiolaetal2015, Cai2017, Kovacs2018, Kovacs2019,  Kovacs2022ISW_eBOSS}, which may indicate a tension with the standard model, but agreement with the standard model was found in \citet{Nadathur2016b} when a matched filter approach was used in the analysis. Using a larger survey area from the DESI legacy survey \citet{Hang2021, Dong2021a, Dong2021} also do not find evidence of an excess ISW signal around voids and superclusters. It has also been argued that through the ISW effect, the Eridanus Supervoid \citep{Szapudi2015} contributes substantially (or modestly; see \citet{Kovacs2022DES_ColdSpot}) to the somewhat anomalous temperature decrement in the Cold Spot \citep{Cruz2005}, a several-degree patch on the CMB.

While the amplitudes of the ISW signal around voids and superclusters may still be debatable among different studies, all the above analyzes have found the sign of the signal to be consistent with expectation from the standard model. These results, together with the positive cross-correlation between the galaxy samples with the CMB, indicate that the late-time gravitational potentials are indeed decaying. This has placed strong constraints on the Galileon model \citep{Renk2017}.

Both the effects of CMB lensing and ISW are gravitational interactions between voids and CMB photons. Other astrophysical processes, such as the thermal Sunyaev-Zel'dovich effect \citep{SZ1970, Sunyaev1980a} (tSZ) also occur between voids and the CMB.

The tSZ effect arises from inverse Compton scattering of free electrons to CMB photons, inducing additional fluctuations for the observed CMB temperature \citep{SZ1970}:
\begin{equation}
\frac{\Delta T}{T_{\rm CMB}} \propto \int n_e T_e d\ell,
\end{equation}
where $n_e$ and $T_e$ are the electron number density and temperature. $d\ell$ integrates along the line of sight between the observer to the CMB. The tSZ effect can be used to measure the projected electron pressure $n_eT_e$ around voids. With additional information about the gas temperature $T_e$, the abundance of baryons, which can be approximated by the number density of free electrons, can be constrained. Detections of the tSZ effect around voids have been reported in \citet{Alonso2018, Li2024}. Measurements of the baryon content in voids, compared with the baryon fractions in other types of the cosmic web: galaxy clusters, filaments and walls, may be able to help to understand the baryon cycle in the Universe.

\section{The impact of a local void on the perception of cosmic expansion}
One intriguing hypothesis is that the observed late-time acceleration of the Universe could be explained by our position near the center of a large, shallow void perhaps hundreds of Mpc across. This would arguably be large enough to challenge the cosmological principle (that we live in a typical region of the Universe). The density of the void increases with the radial distance $R$ from the observer. Eventually, at sufficiently large $R$, the average density of the void approaches the mean density of the Universe. Observers within this void would perceive a locally faster expansion rate compared to a homogeneously matter-dominated Universe. As $R$ increases and the density contrast diminishes, the observed expansion rate would increasingly resemble that of a standard matter-dominated Universe. By adjusting the density profile of such a void, it is possible to mimic the observed accelerated expansion history of the Universe in a matter-dominated universe without dark energy \citep[e.g.][]{Barausse2005, Alnes2006, Garcia-Bellido2008}.

However, the hypothesis was largely ruled out by observations of the kinetic Sunyaev-Zeldovich (kSZ) effect \citep{Sunyaev1980a, Sunyaev1980}, which measures the peculiar velocities of galaxy clusters through the CMB rest frame. 
If the local void is large and deep enough to explain dark energy, we would expect a much larger kSZ effect than we have actually observed. The observed kSZ effect is in fact consistent with the predictions of the $\Lambda$CDM model and this places stringent constraints on large-scale inhomogeneities, effectively disfavoring the void model as a primary explanation for cosmic acceleration \citep{Zhang2011}.

There is, however, observational evidence that our nearby environment is slightly underdense. Observations of galaxy number counts and the luminosity function suggest a local underdensity \citep[e.g.][]{Keenan2013, Whitbourn2014}, which could lead to the perception that the local Universe is expanding slightly faster than elsewhere. However, it is disputed whether this is sufficient to fully explain the Hubble tension -- the apparently larger value of the Hubble constant measured locally compared to that inferred from CMB observations in the $\Lambda$CDM model \YC{\citep[e.g.][]{Shanks2019, Kenworthy2019, Cai2021,Castello2022, Mazurenko2025}.}

This local void is not to be confused with the Local Void, which has been clearly mapped. The Local Sheet, a cosmic wall in which the Milky Way is embedded, forms a boundary of the Local Void, a perhaps remarkably \citep{Peebles2010} deep void some 60 Mpc across \citep{Tully1987,Tully2019}.

\section{Summary}

Cosmic voids as part of the cosmic web are some of the largest-scale structures of our Universe. The low-density nature of voids makes them unique in preserving the initial conditions of the Universe, and having arguably more direct sensitivity to the global expansion of the Universe. Analysing the large-scale structure from the perspective of cosmic voids can often provide us with new and complementary information about cosmology and astrophysics. They are useful in constraining cosmological parameters including dark energy, measuring the sum of neutrino masses and testing theories of gravity. 

As a cosmological probe, cosmic voids have now been established as one of the beyond-two-point statistics. When combined with the conventional two-point statistics, they can often provide extra cosmological information. In the era of precision cosmology where systematic errors are dominating our measurements, this is both valuable and challenging. The reduction of statistical errors brought by extra probes from voids is likely to increase the requirement for the control of observational and theoretical systematics. Next generation galaxy surveys such as the ongoing surveys of DESI, LSST and Euclid promise to increase the current survey volume and number of galaxies by an order of magnitude. This offers great opportunities for advancing our understanding of cosmology using multiple cosmological probes. However, calibration of observational systematics is likely to remain the major challenge. Perhaps more valuable is to analyze the data independently, from the perspective of only cosmic voids. Being different from the conventional analyzes, this may help to provide diagnostics for observational systematics.

\mn{Finally, for further information on voids, we note that there are a few nice review articles written for researchers \citep{vandeWeygaert2011,vdW2016, Cai2018, Pisani2019}. }

\begin{ack}[Acknowledgments]
\YC{We thank Rien van de Weygaert for useful comments.} YC thanks John Peacock for useful discussions during the preparation of this manuscript. YC acknowledges the support of the Royal Society through a University Research Fellowship. For the purpose of open access, the author has applied a Creative Commons Attribution (CC BY) license to any Author Accepted Manuscript version arising from this submission.
\end{ack}

\seealso{Cosmic web \citep{Tojeiro2025}}

\bibliographystyle{Harvard}
\input{Void_V9.bbl}


\end{document}